\documentclass[letterpaper, 10 pt, conference, onecolumn]{ieeetran}
\IEEEoverridecommandlockouts
\usepackage[utf8]{inputenc}

\usepackage{graphicx}
\usepackage{amsmath}
\usepackage[version=4]{mhchem}
\usepackage{longtable,tabularx}
\usepackage{nomencl}
\makenomenclature

\usepackage{amsmath,amssymb,amsfonts}
\usepackage[a-2b,mathxmp]{pdfx}[2018/12/22]
\usepackage{algorithmic}
\usepackage{graphicx}
\usepackage{textcomp}
\usepackage{caption}
\usepackage{subcaption}
\usepackage{tabularx}
\usepackage{xcolor}
\usepackage{booktabs}
\usepackage{multicol}
\usepackage{multirow}
\usepackage{mathtools}

\usepackage{etoolbox}
\usepackage{physics}
\usepackage{hyperref}
\usepackage{fancyhdr}
\fancypagestyle{specialfooter}{%
  \fancyhf{}
  
  \fancyfoot[L]{\footnotesize This work was presented at the AIAA Aviation 2023 Forum. \\ Copyright @ Souma Chowdhury}
  \fancyfoot[C]{1}
}
\usepackage{kantlipsum}
\usepackage[style=ieee,maxbibnames=99]{biblatex}
\makeatletter
\newcommand{\linebreakand}{%
  \end{@IEEEauthorhalign}
  \hfill\mbox{}\par
  \mbox{}\hfill\begin{@IEEEauthorhalign}
}
\makeatother
\newcommand{\G}{\mathcal{G}}
\newcommand{\Ss}{\mathcal{S}}
\makeatletter
\patchcmd{\@makecaption}
  {\\}
  {.\ }
  {}
  {}
\makeatother
\def\tablename{Table}
\begin{document}

\title{ Physics-Infused Machine Learning Based Prediction of VTOL Aerodynamics with Sparse Datasets}
\author{Manaswin Oddiraju \footnote{Ph.D. Student, Department of Mechanical and Aerospace Engineering, University at Buffalo, AIAA Student member.}, Divyang Amin \footnote{Flight Sciences Engineering Lead, Bechamo LLC}, Michael Piedmonte\footnote{Chief Technical Officer, Bechamo LLC}, Souma Chowdhury\footnote{Associate Professor, Department of Mechanical and Aerospace Engineering,University at Buffalo, AIAA Senior member. Corresponding author. Email: soumacho@buffalo.edu}}

\maketitle
\thispagestyle{plain}
\pagestyle{plain}

\thispagestyle{specialfooter}

\begin{abstract}
\it
    Complex optimal design and control processes often require repeated evaluations of expensive objective functions and consist of large design spaces. Data-driven surrogate models such as neural networks and Gaussian processes provide an attractive alternative to expensive simulations and are utilized frequently to represent these objective functions in optimization. However, pure data-driven models, due to a lack of adherence to basic physics laws and constraints, are often poor at generalizing and extrapolating. This is particularly the case, when training occurs over sparse high-fidelity datasets. A class of Physics-infused machine learning (PIML) models integrate ML models with low-fidelity partial physics models to improve generalization performance while retaining computational efficiency. This paper presents two potential approaches for Physics infused modelling of aircraft aerodynamics which incorporate Artificial Neural Networks with a low-fidelity Vortex Lattice Method model with blown wing effects (BLOFI) to improve prediction performance while also keeping the computational cost tractable. This paper also develops an end-to-end auto differentiable open-source framework that enables efficient training of such hybrid models. These two PIML modelling approaches are then used to predict the aerodynamic coefficients of a 6 rotor eVTOL aircraft given its control parameters and flight conditions. The models are trained on a sparse high-fidelity dataset generated using a CHARM model. The trained models are then compared against the vanilla low-fidelity model and a standard pure data-driven ANN. Our results show that one of the proposed architecture outperforms all the other models and at a nominal increase in execution time. These results are promising and pave way to PIML frameworks which are able generalize over different aircraft and configurations thereby significantly reducing costs of design and control.
\end{abstract}

\printnomenclature

\section{Introduction}

Real world optimal design and control problems consist of optimizations with large design spaces and computationally expensive objective functions and therefore require efficient and accurate models of complex physical systems. There are different proposed approaches to render these problems computationally feasible, such as the use of surrogate models, reduced order modelling methods and multi-fidelity optimization techniques. The use of data-driven surrogate models for optimization \cite{chatterjee2019critical} is a popular approach to reduce the computational burden and render the process tractable. However, using pure data driven models poses additional challenges in the form of poor explainability and usually these models require large amounts of data either from expensive high-fidelity analysis or experiments in order to achieve sufficient accuracy. For many engineering design problems however fast running physics-based models are available, but they usually trade-off accuracy for performance.  Therefore, in this paper, we propose two Physics Informed Machine Learning (PIML) models which combine data driven Artificial Neural Networks (ANNs) with low-fidelity physics models and augment the performance of the physics models while only needing a sparse high-fidelity dataset. We also develop an end-to-end auto-differentiable physics framework to enable efficient training of such hybrid architectures and apply them to model the aerodynamics of a 6 rotor eVTOL aircraft. The rest of this section briefly covers existing techniques for augmenting low-fidelity physics models using high-fidelity data, physics informed machine learning architectures before stating our research objectives.

Conventional methods for enhancing low-fidelity computational models with high-fidelity or experimental data encompass techniques like Bayesian Inference \cite{bisaillon2022combined, morelli1997accuracy}, Gradient-Based and Gradient-Free Optimization \cite{mohamad2020dynamic, morelli2005aerodynamic}, among others. The primary goal of these techniques is to determine the optimal parameters of the low-fidelity model in order to improve its overall performance. However, due to the complexity of the physical phenomena often modeled by these techniques, identifying a static set of parameters for a dataset might be inadequate and may not fully capitalize on the potential accuracy improvements offered by a sparse high-fidelity dataset. On the other hand, data-driven models, due to their near-instantaneous run times, are being utilized frequently for speculating the behavior of complex systems in domains such as mechanical systems \cite{bataineh2016neural,garimella2017neural}, robotics \cite{pfeiffer2017perception,shao2019unigrasp,rajeswaran2017learning}, and energy forecasting~\cite{ghassemi2017optimal}. Models such as Neural Networks, Gaussian Processes etc. are showing competitive prediction accuracy on some problems in this domain. However, they underperform at extrapolating and generalizing \cite{haley1992extrapolation,neal2012bayesian}, especially when trained with small or sparse data sets \cite{solomatine2009data}. This can be attributed to a lack of adherence to basic laws of physics. Additionally, they also exhibit challenging-to-interpret black-box behavior and sensitivity to noisy data \cite{karimi2020deep}. In lieu of these problems, significant research efforts are being directed towards "hybrid models" or "physics-infused" models. These models generally integrate data-driven and low-fidelity physics models in order to make predictions computationally feasible and accurate.

There are different classes of  physics-infused neural networks; although almost all of them make use of a data-driven model and a physics model, they differ in the architectures. Hybrid ML architectures can broadly be classified into  serial\cite{narendra1990identification,young2017physically,nourani2009combined,singh2019pi} and parallel \cite{javed2014robust,cheng2009fusion,karpatne2017physics} architectures. Serial architectures typically have the data-driven models set in sequence with the partial physics model or used to tune the partial physics model parameters, while parallel architectures usually contain additive or multiplicative ensembles of partial physics and data-driven ML models \cite{mangili2013development}. Several such hybrid PIML architectures have been reported in the literature in the past few years \cite{kapusuzoglu2020physics,rufa2020towards,rai2021hybrid,matei2021controlling,zhang428midphynet,chen2021hybrid,choi2021hybrid,machalek2022dynamic,lai2022intelligent,freeman2022physics,rajagopal2022physics,ankobea2022hybrid,maier2022known}, spanning over a wide range of applications such as in modeling dynamic systems, cyber-physical systems, robotic systems, flow systems and materials behavior, among others.
The OPTMA model \cite{behjat2020physics}, is a physics-infused machine learning model which combines an artificial neural network with a partial physics model in order to make predictions. Earlier applications of this framework on acoustics problems have shown that it generalizes well and can even extrapolate successfully. 
This architecture primarily works by using the data-driven model -in this case the transfer network;  to map the original inputs to the inputs of the partial physics model so as to make its output match that of the high-fidelity model. Hence, the training process is just the transfer network learning this mapping. And since the intermediate parameters are open and interpretable to domain experts (as they are inputs to partial physics models), the OPTMA model is less of a black box and more understandable when compared to pure data-driven models.

In all of these sequential hybrid-ML models however, the presence of an external partial physics model increases the complexity and cost of the training process. Previously\cite{iqbal2022efficient}, this hurdle was overcome by programming custom loss functions which include the partial physics in PyTorch\cite{NEURIPS2019_9015} to enable backpropagation. However, this approach may not be feasible in all domains as PyTorch is not optimal for general purpose scientific computing (especially numerical methods). Therefore, in our goal to make OPTMA more user-friendly and computationally tractable, we use Google JAX \cite{jax2018github} to auto-differentiate the partial physics model. JAX is capable of forward and backward mode auto-differentiation and works on codes written in python and is also easy to integrate with the transfer network written in PyTorch. This is a much more generalizable and scalable approach to creating computationally efficient PIML frameworks. 

During the process of aircraft design, and eventually the aircraft controller design, engineers often encounter the earlier stated predicament of selecting appropriate methods to model the aircraft. They may need to balance the benefits and drawbacks of using a method that is highly accurate, albeit costly and time-consuming, against a faster and more economical approach that might not provide the same level of precision.The complexity of this problem is magnified when engineers are tasked with designing more challenging-to-control aircraft than traditional fixed-wing varieties. A prime example is the Vertical Takeoff \& Landing (VTOL) aircraft. As the aircraft designs become more complex, the modeling demands can escalate exponentially. In the case of VTOL, it is essential to model the aircraft dynamics across all phases: from hover to transition, and finally into cruise. Such aircraft also exhibit more intricate aerodynamics, including phenomena like blown-wing effects. There is typically a greater number of degrees of freedom, which further escalates the complexity of designing robust controllers for VTOL aircraft. Given the numerous benefits and strengths of physics-infused modeling, and the necessity for fast accurate models for aircraft design and control, it is evident that PIML models are particularly well-suited for modeling the aerodynamics of an aircraft. 

Therefore, in this paper, we propose two physics informed modelling approaches to augment a low-fidelity VLM model with a sparse high-fidelity dataset. One of the approaches, called PIML-A is an extension of the OPTMA architecture. The other architecture, named PIML-B is a comparatively simpler ensemble model that learns the error in the low-fidelity physics models. The architectures are aimed at improving the accuracy of the low-fidelity physics models while only using a sparse high-fidelity dataset to keep training costs low. Our research objectives are as follows: 
\begin{enumerate}
    \item Create a Vortex Lattice Method (VLM) model with auto-differentiation capabilities.
    \item Develop different PIML architectures seeking to provide better accuracy / cost trade-offs than pure Low-Fidelity and pure data-driven models.
    \item Apply the above architectures to model the control inputs to force and moment coefficients of a eVTOL aircraft.
    \item Compare the performance and computational cost of the PIML models with the baseline low-fidelity and data-driven models. 
\end{enumerate}

The remainder of this paper is laid out as follows: Section\ref{sec:framework} contains a description of the two PIML models framework and their components in detail. Section\ref{sec:casestudy} explains our case study and the models that we are using before moving on to Section\ref{sec:results} which contains our results and discussion. Finally, we state our conclusions in Section\ref{sec:conclusion}. 

\section{Physics Infused Machine Learning Architectures}
\label{sec:framework}

\begin{figure}[!h]
\begin{subfigure}{\linewidth}
    \centering
    \includegraphics[width=\linewidth]{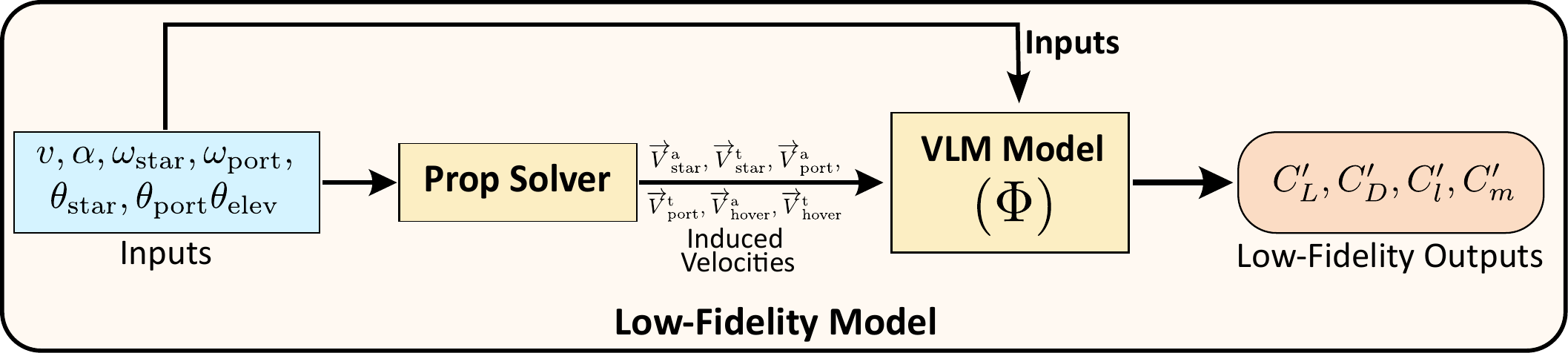}
    \caption{}
    \label{fig:lf_model}
\end{subfigure} 

\begin{subfigure}{\linewidth}
    \centering
    \includegraphics[width=\linewidth]{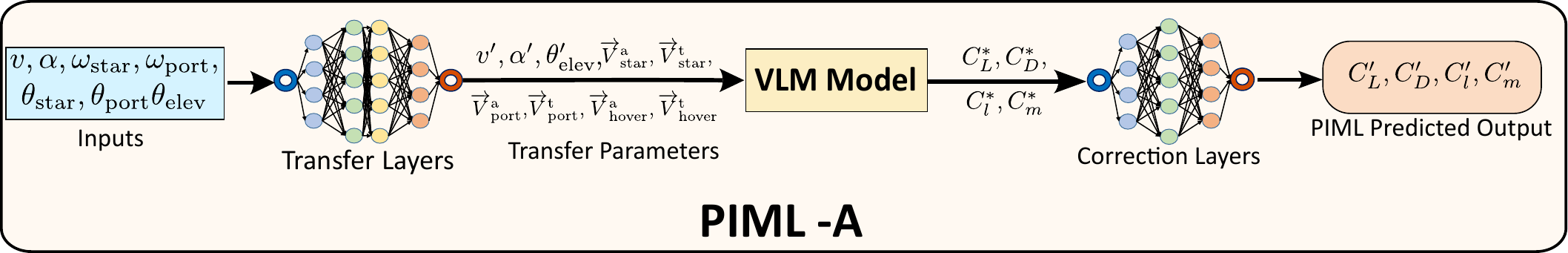}
    \caption{}
    \label{fig:PIML_A}
\end{subfigure}  

\begin{subfigure}{\linewidth}
    \centering
    \includegraphics[width=0.9\linewidth]{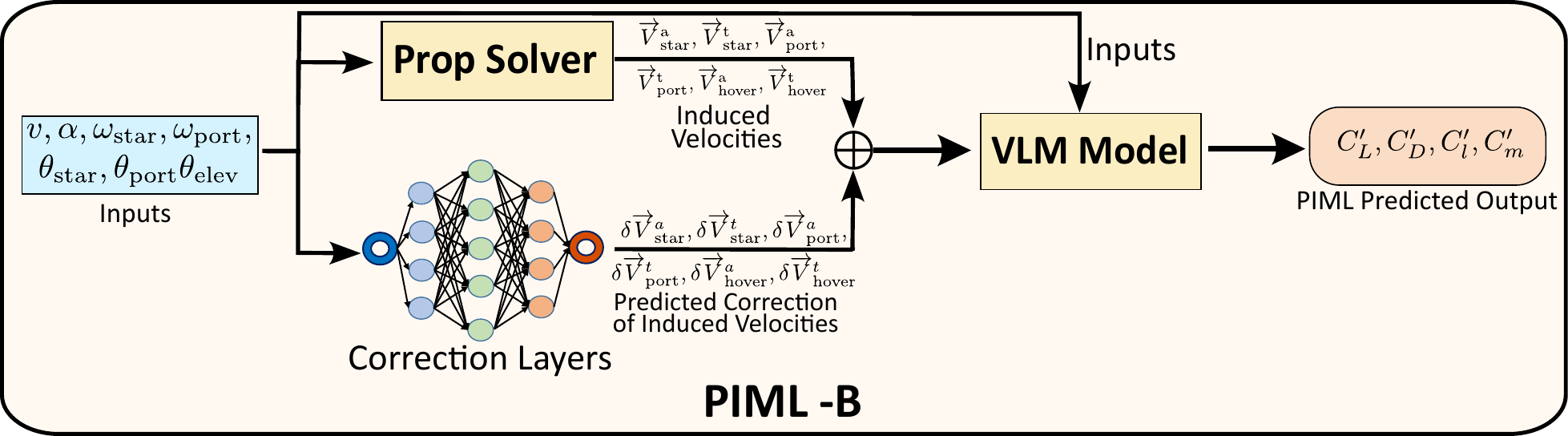}
    \caption{}
    \label{fig:PIML_B}
\end{subfigure} 

\begin{subfigure}{\linewidth}
    \centering
    \includegraphics[width=0.8\linewidth]{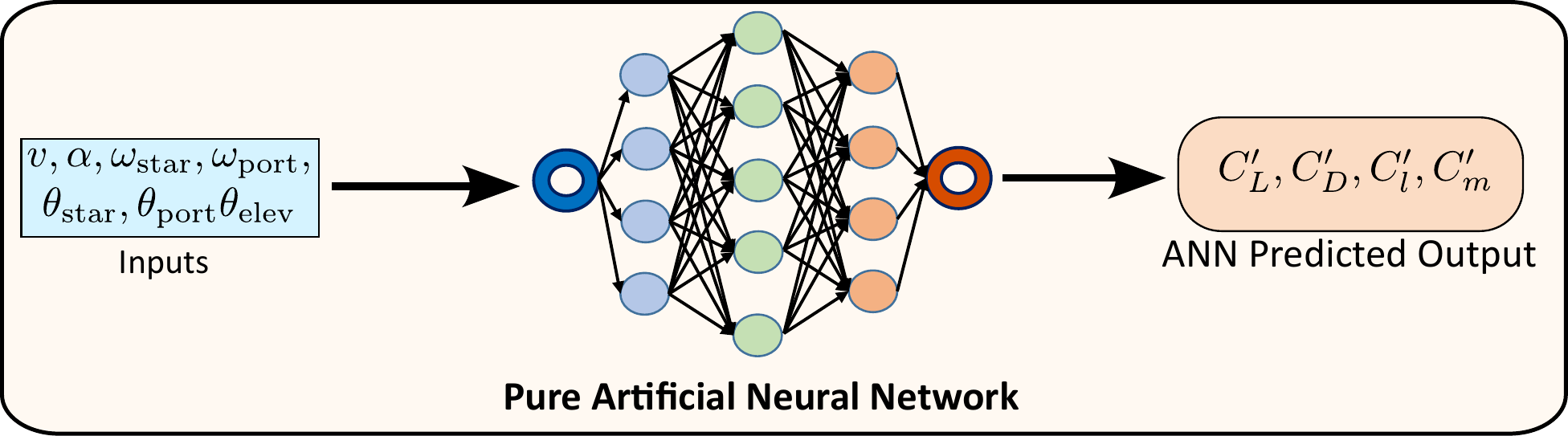}
    \caption{}
    \label{fig:ANN}
\end{subfigure} 
\caption{Framework Diagrams of a) Low-Fidelity Physics b) PIML-A  c) PIML-B and d) Purely Data-Driven ANN }
\label{fig:framework}
\end{figure}

Fig.\ref{fig:framework} shows the two proposed PIML frameworks as well as the vanilla low-fidelity physics model ($\mathcal{L}$). In this paper, all the models shown are used to predict the airframe force and moment coefficients for a given set of control and environmental parameters. The PIML models amalgamate low-fidelity physics models with Artificial Neural Networks (ANNs) to enhance predictions, yet they vary based on the specific role of the ANNs. As depicted in the framework diagram \ref{fig:framework}, the low-fidelity model comprises of a propeller solver and the Vortex Lattice Method (VLM) model. The prop solver computes induced velocities which are then fed into the VLM, which in turn calculates the force and moment coefficients on the airframe.

The PIML-A model incorporates ANN layers to achieve three primary objectives: 1) To supersede the prop solver and directly estimate the axial and tangential induced velocities of the propellers, 2) to modify the flight conditions in order to enhance the predictions of the VLM model, and 3) to adjust the output of the VLM to reduce discrepancies with respect to the high-fidelity data. Given that all layers are trained concurrently, the PIML-A model learns the mapping from flight conditions and control inputs to the transfer parameters. Simultaneously, it learns the output errors of the VLM and adjusts them to align more closely with the high-fidelity data. The reasoning behind this model is that part of the errors in the VLM model can be mitigated by shifting the inputs into different domains, and the errors that remain, not resolved by this input modification, will be addressed by the correction layers on the output side.

In contrast, the PIML-B model does not adjust any inputs to the original low-fidelity model and retains the prop solver in its structure. Within this design, the sole objective is to identify the error in the induced velocities computed by the prop solver and rectify these parameters. This model is more closely aligned with the low-fidelity physics model and aims to enhance modeling accuracy by refining the outputs of the prop solver.

The following subsection focuses on the training processes of these PIML models, specifically the computation of gradients and the implementation of backpropagation through the low-fidelity physics models.
 
\subsection{PIML Model Training}
The training of neural networks (i.e optimizing the weights of the neural network to minimize loss) is performed using gradient descent and therefore requires the gradient of the loss function($L$) w.r.t to the weights($W$). The following equations show the integration of backpropagation gradients through an auto-differentiable physics model. This feature enables the training of our PIML architectures using standard ANN training techniques and using standard ANN software libraries.

\begin{equation}
    \begin{aligned}
        L=& \left\|Y_i-Y_i^{\prime}\right \|^2 \text{(MSE Loss)} \\ 
        L=&\frac{1}{n} \sum_{i=1}^n\left(Y_i-Y_i^{\prime}\right)^2 \\
        \therefore \quad \frac{\partial L}{\partial W}=& \frac{1}{n}\sum_{i=0}^n 2\left(Y_i^{\prime}-Y_i\right)\frac{\partial Y_i^{\prime}}{\partial W}\\
        \text{Where } \underbrace{Y_i}_{\text{Ground Truth}} &= \phi(X_i)\text{,  }\underbrace{Y_i^{\prime}}_{\text{Prediction}} = \G(X_i,W) 
    \end{aligned}
\end{equation}
$\G$ represents the sequential PIML model and $\phi$ represents the high-fidelity physics model.The terms $X_i$, $Y_i$, $Y_i^\prime$ represent the input, high-fidelity output and the OPTMA prediction of the $i^{\rm{th}}$ training sample respectively. \\
From the definition of $Y_i ^ {\prime}$ :
\begin{equation}
    \begin{aligned}
        Y_i^{\prime}=& \G\left(X_i, W\right) \\ 
        \frac{\partial Y_i^{\prime}}{\partial W}= & \frac{\partial G\left(X_i, W\right)}{\partial W} \\ 
         \text{As } \G \text{ is the OPTMA model, we have:}  \\ 
        \G\left(X_i, W\right) &=\Phi\left(\Ss\left(X_i, W\right)\right) \\
        \implies \Aboxed{\frac{\partial \G\left(X_i, W\right)}{\partial W} =\frac{\partial \Phi\left(\Ss \left(X_i, W\right)\right)}{\partial W} &=\underbrace{\frac{\partial \Phi(U)}{\partial U}}_{\substack{\text{Partial Physics}\\ \text{ Auto-Differentiation}}}\times \underbrace{ \frac{\partial \Ss \left(X_i, W\right)}{\partial W}}_{\rm{Backpropagation}}} \\
    \end{aligned}
\end{equation}
Here, $\Ss$ and $\Phi$ represent the neural network and the partial physics models respectively. $U$ represents the intermediate parameter vector which is the output of the transfer network and input to the partial physics model.

\section{Case Study: Modelling Aerodynamics of A Tilt-Rotor eVTOL Aircraft }
\label{sec:casestudy}
The Aerobeacon, shown in fig.\ref{fig:aircraft2}, is a 6 rotor aircraft with 4 hover rotors in a quad configuration and two tilt-rotors at the edges of the wings. The aircraft has no rudder and instead makes use of differential thrust between the tilt rotors for yaw control. The only movable control surface on the aircraft apart from the tiltrotor is an elevator. Since this is our initial attempt at modelling this aircraft, we chose to fix the hover rotor RPMs at 5000 and only vary the other control inputs listed in Tale.\ref{tab:bounds}.

\begin{minipage}{0.39\linewidth}
    \small
    \captionof{table}{Aircraft Control Parameter Bounds}
    \label{tab:bounds}
    \centering
    \begin{tabularx}{\linewidth}{>{\centering}X >{\centering}X}
    \toprule
    Parameters & Bounds \tabularnewline
    \midrule
     Speed ($v$)   &  [0 m/s ,45 m/s]  \tabularnewline
     Angle of Attack ($\alpha$)& [-15$^{\circ}$ , 15$^{\circ}$ ] \tabularnewline
     Starboard Propeller RPM ($\omega_{\rm{star}}$)& [4000 , 10000] \tabularnewline
     Port Propeller RPM ($\omega_{\rm{port}}$) & [4000 , 10000] \tabularnewline
     Starboard Propeller Angle ($\theta_{\rm{star}}$) & [0$^{\circ}$  , 110$^{\circ}$ ]  \tabularnewline
     Port Propeller Angle ($\theta_{\rm{port}}$) & [0$^{\circ}$  , 110$^{\circ}$ ]  \tabularnewline
     Elevator Deflection ($\theta_{\rm{elev}}$) & [-15$^{\circ}$ , 15$^{\circ}$ ] \tabularnewline
     \bottomrule
    \end{tabularx}

\end{minipage} \hfill
\begin{minipage}{0.55\linewidth}
    \centering
    \includegraphics[width=.9\linewidth]{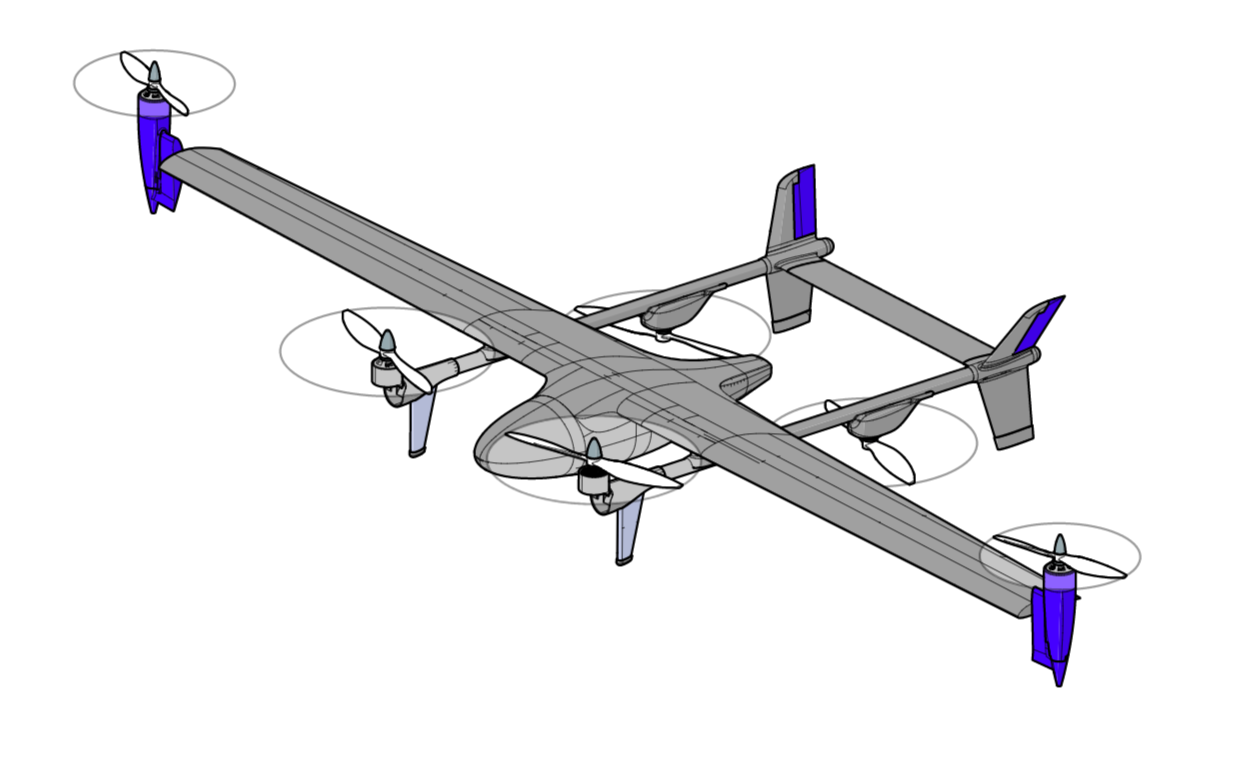}
    \captionof{figure}{Aerobeacon aircraft design configuration from Flighthouse Engineering}
    \label{fig:aircraft2}
\end{minipage}


\subsection{Modelling Aircraft Flight Dynamics}
Table \ref{tab:modelling} displays the inputs and outputs of all the models and also the transfer parameters corresponding to the PIML-A model. The transfer parameters are specifically selected such that they have an impact on the output accuracy of the model, as well as to allow greater flexibility to the embedded ANN layers to affect the final outputs. In both the proposed architectures, we try to improve the low-fidelity model by either augmenting or replacing the prop solver. The PIML-A model in particular also tries to alter the outputs of the VLM model to improve accuracy. This model is an extension of the OPTMA architecture discussed in the Introduction.

\begin{table}[hp]
    \caption{Modeling Parameters}
    \label{tab:modelling}
    \centering
    \small
    \begin{tabularx}{\linewidth}{>{\centering}m{0.2\linewidth} >{\centering}X >{\centering}X}

    \toprule 
    \hline
     Inputs (All Models)  & Transfer Parameters (PIML-A Model) &  Outputs (All Models) \tabularnewline
    \midrule 
      Speed ($v$)\newline  Angle of Attack ($\alpha$)\newline  Starboard Propeller RPM ($\omega_{\rm{star}}$)\newline Port Propeller RPM ($\omega_{\rm{port}}$)\newline  Starboard Propeller Angle ($\theta_{\rm{star}}$)\newline  Port Propeller Angle ($\theta_{\rm{port}}$)\newline  Elevator Deflection ($\theta_{\rm{elev}}$) & 
  $\begin{array}{c} \rm{Speed} (v^\prime)\\ \text{Angle of Attack} (\alpha^\prime)\\  \text{Elevator Deflection} (\theta_{\rm{elev}}^\prime) \\
       \overrightarrow{v}_{\rm{star}}^a  \\  \overrightarrow{v}_{\rm{star}}^t  \\  \overrightarrow{v}_{\rm{port}}^a  \\  \overrightarrow{v}_{\rm{port}}^t  \\ \overrightarrow{v}_{\rm{hover}}^a \\ \overrightarrow{v}_{\rm{hover}}^t \end{array}$ & $ \begin{array}{c} C_L \\ C_D \\C_l \\ C_m \end{array}$ \tabularnewline
     \hline
     \bottomrule
    \end{tabularx}
\end{table}

\subsection{Low Fidelity Model}
The lower fidelity flight dynamics model uses Bechamo LLC's developed VLM plus propeller effects low-fidelity tool, BLOFI.  This tool was developed initially to model the NASA Tiltwing aircraft and was validated against the X-57 data \cite{Duensing_Housman_Maldonado_Kiris_Yoo_2019}. The validation plots can be seen in fig.\ref{fig:X-57 vaidation}. This was done in order to prove that we are able to capture the blown wing effects. BLOFI was created by using an improvised VLM that accounts for some blown wing effects caused by the propellers upstream of the the lifting surfaces. A base VLM \cite{Drela2014Aerodynamics} is implemented and is modified by accounting for induced velocities caused by propellers' wash upstream of the control points on the lifting surfaces. The properties for each propeller are calculated using Blade Element Momentum Theory \cite{QPROP}. It should also be noted that propeller slipstream contraction \cite{Veldhuis_2028} is accounted for when including the induced velocities from the propellers in the VLM setup.

A set of flight conditions is created by making a grid of every combination of freestream velocity, sideslip and angle of attack. At each of these flight conditions the aerodynamic forces and moments, including the propeller thrusts and torques, are calculated, using the aforementioned VLM implementation. This is done for every expected RPM setting, propeller tilt angle and tail deflection. The combination of aerodynamic forces and moments and propeller thrusts and torques is used to determine the overall forces and moments acting on the aircraft.

\begin{figure}[b]
\begin{minipage}{0.45\linewidth}
    \centering
     \includegraphics[width=.9\linewidth]{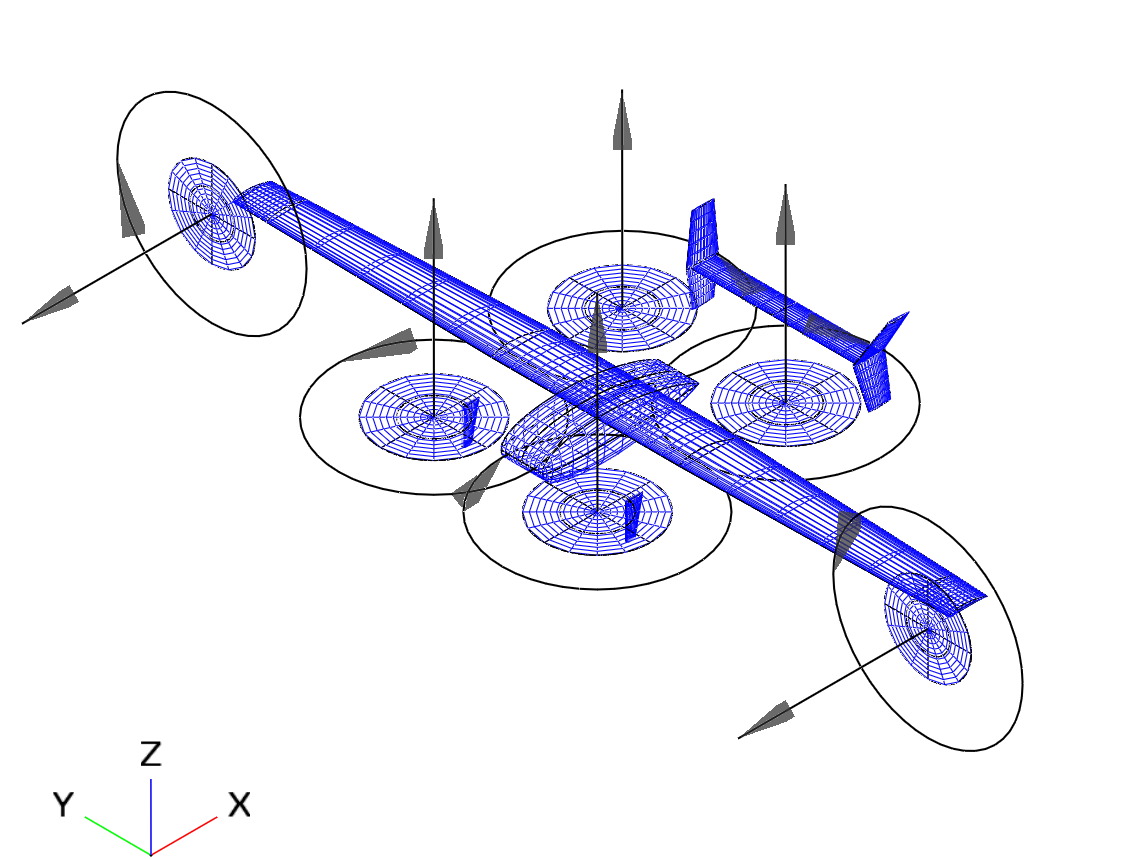}
    \captionof{figure}{Mesh model of the Aerobeacon aircraft. The propellers and their direction of rotation are represented by directed circles. The two rotors positioned at the wing extremities are capable of tilting. }
    \label{fig:aircraft}
\end{minipage} \hfill
\begin{minipage}{0.45\linewidth}
    \centering
    \includegraphics[width=0.9\linewidth]{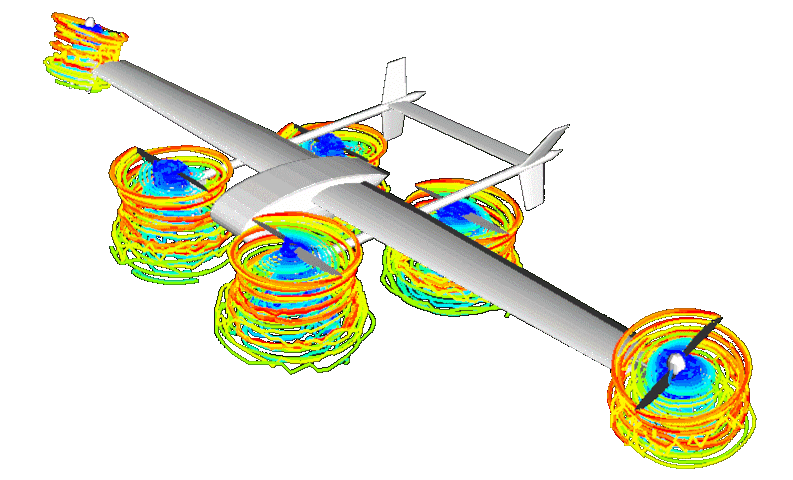}
    \captionof{figure}{CHARM visualized output of particle method to obtain a higher-fidelity model at discrete points}
    \label{fig:still_charm}
\end{minipage}
\end{figure}

CHARM\cite{CHARM} is a commercial tool developed by Continuum Dynamics, Inc. for simulating the aerodynamics and dynamics of rotorcraft using partical methods. Flighthouse Engineering LLC. provided their Aerobeacon aircraft design, parameters, and constructed the CHARM model.  They ran CHARM test cases for a collection of randomly generated input samples. The data set derived from the CHARM model serves as the high-fidelity data for the research presented in this paper.
The VLM configuration is shown in fig.\ref{fig:aircraft} and the CHARM higher fidelity output visualization is shown in fig.\ref{fig:still_charm}.

\subsection{Sparse High-Fidelity Dataset}

In order to generate the high fidelity dataset, we used Latin Hypercube Sampling and created a dataset of 100 points across the parameters and bounds listed in Tab.\ref{tab:bounds}. These 100 samples were then input to the CHARM high fidelity model and the dataset was generated. Out of the 100 initial samples, only 87 were usable as 11 cases failed to converge on the CHARM model and the remaining 2 cases showcased abnormally high values of the force and moment coefficients and were removed from the dataset. While generating the dataset, we fixed the RPMs of the hover quad rotors to 5000 RPM. 
This final dataset of 87 samples was then randomly divided into a training set containing 70 samples and a validation set containing 17 samples. All the performance results shown in this paper are run on this validation dataset consisting of samples completely unseen by the model.

\subsection{Prior Analysis of BLOFI Performance on Other Data-Sets}
BLOFI was developed at Bechamo LLC and in order to prove its usefulness in comparison to other available VLM solvers that do not account for blown wing effects. It has been previously validated using CFD data\cite{Duensing_Housman_Maldonado_Kiris_Yoo_2019} generated by NASA for the X-57. Just as a flat plate model was created for the Aerobeacon with the appropriate propellers for the developments of this paper, a flat plate model with the corresponding propellers was earlier created for the X-57 as part of the validation study on BLOFI. In figure\ref{fig:X-57 vaidation} we are able to see that BLOFI can capture significant blown-wing effects, by comparing the power-on and power-off plots. The increase in $C_L$ and $C_D$ at every angle of attack can be observed. More importantly, the inclusion of induced velocities in the VLM solve also lets us observe the increased flap effectiveness. Errors are between 5$\%$ and 10$\%$ for the range of angles of attack for which this validation study was conducted. It is expected that with a PIML approach, not only can we close down this gap (error), but more readily extend the low/medium fidelity solver such as BLOFI to work on a wider variety of aircraft configurations with little to no manual tuning and manual model development. To support his premise, the next section presents the results of our PIML architectures with the example of Aerobeacon configuration, where we had more control over the high-fidelity sampling -- enabling ease of PIML training for this paper. 


\begin{table}[]
    \small
    \caption{Comparison of PIML, Low-Fidelity and Pure ANN Models }
    \label{tab:ANN_params}
    \centering
    \begin{tabularx}{\linewidth}{>{\raggedright}p{0.17\linewidth} >{\centering}p{0.22\linewidth} >{\centering}p{0.22\linewidth} >{\centering}p{0.22\linewidth} >{\centering}X}
    \toprule
    \hline
                              & PIML-A  &  PIML-B  &  Low-Fidelity & ANN \tabularnewline
    \midrule 
        Execution Time (s)    & 0.32  & 0.53    &  0.338    &  0.0002   \tabularnewline 
        Training Time (min)   & 23.46    & 23.18      & -    & 0.04  \tabularnewline  
        $\#$ Hidden Layers    &    6    &  6       & -       & 4  \tabularnewline 
        $\#$ Nodes per Layer  &    200  &  200     & -    & 150 \tabularnewline 
        \hline
    \bottomrule
    \end{tabularx}
\end{table}

\section{Results and Discussion}
\label{sec:results}
\begin{figure}
\centering
    \begin{subfigure}{0.49\linewidth}
        \includegraphics[width = \linewidth]{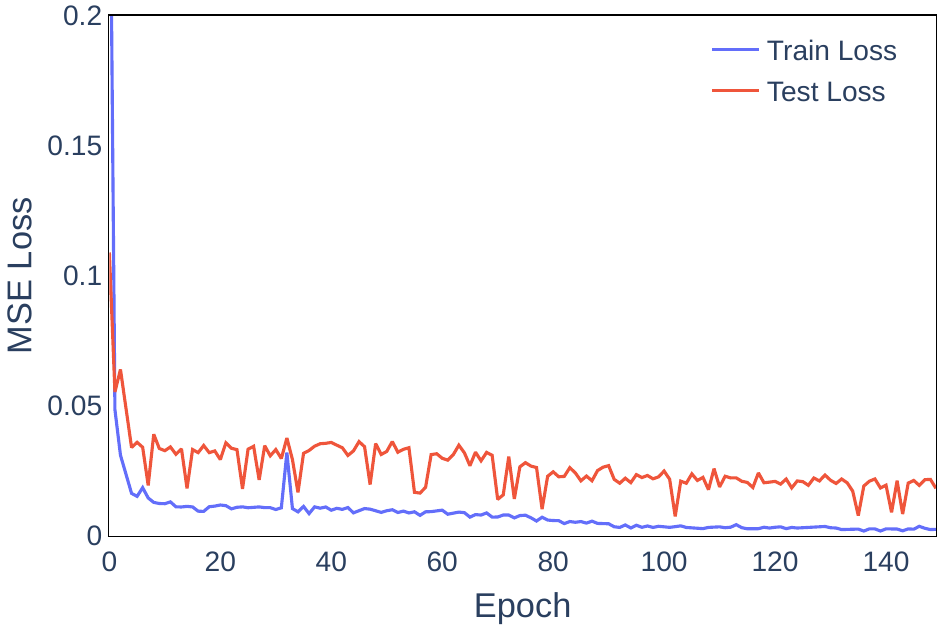}
        \caption{PIML-A}
    \end{subfigure} \hfill 
    \begin{subfigure}{0.49\linewidth}
        \includegraphics[width = \linewidth]{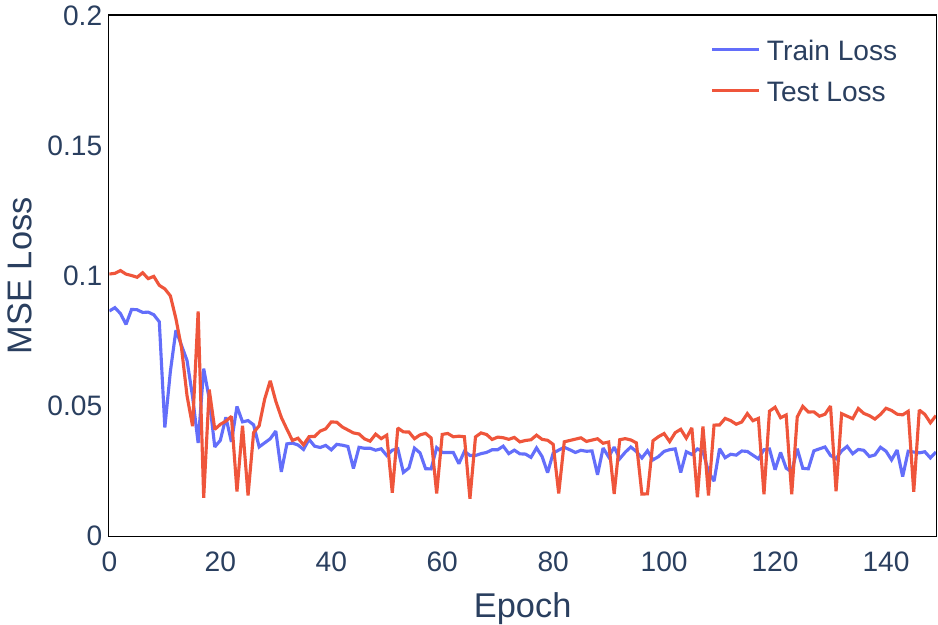}
        \caption{PIML-B}
    \end{subfigure}
    \caption{Convergence history of PIML Models}
    \label{fig:ConvHist}
\end{figure}

\subsection{Baselines For Comparison }
For testing the performance of the two proposed PIML models, we compared them against the vanilla low-fidelity model($\mathcal{L}$) shown in fig.\ref{fig:lf_model} and a pure data-driven ANN ($\mathcal{A}$). Table \ref{tab:ANN_params} lists the parameters of the stand-alone ANN and the ANNs used as part of the PIML frameworks. All the data-driven and PIML models were trained on the same training dataset of 70 high-fidelity samples and validated on a validation dataset consisting of 17 samples.
As the training data was sparse, a pure ANN with similar size of the ANNs used for the PIML models tended to overfit very quickly. Therefore, the size of the pure data-driven ANN is lower as compared to the ANNs used in the PIML models.

\begin{figure}[!h]
    \centering
    \begin{subfigure}{0.49\linewidth}
        \includegraphics[width = \linewidth]{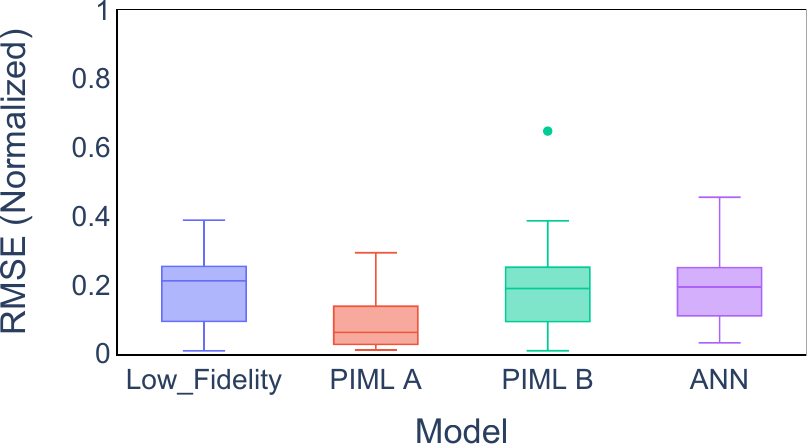}
        \caption{Prediction Error in $C_L$}
    \end{subfigure} \hfill 
    \begin{subfigure}{0.49\linewidth}
        \includegraphics[width = \linewidth]{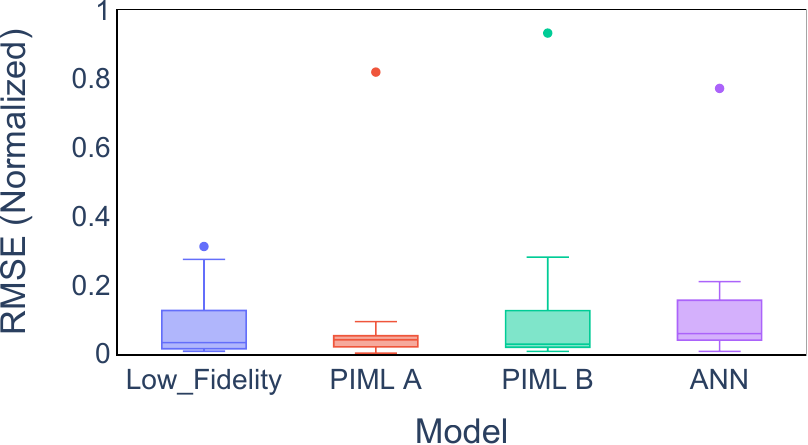}
        \caption{Prediction Error in $C_D$}
    \end{subfigure} \\ 
    \begin{subfigure}{0.49\linewidth}
        \includegraphics[width = \linewidth]{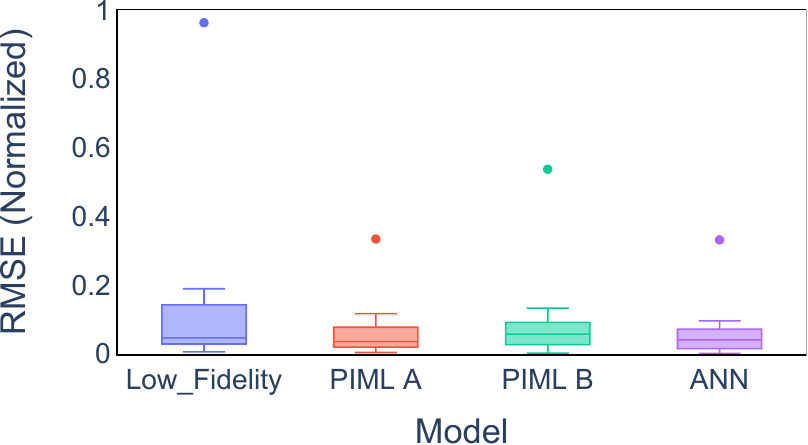}
        \caption{Prediction Error in $C_l$}
    \end{subfigure} \hfill 
    \begin{subfigure}{0.49\linewidth}
        \includegraphics[width = \linewidth]{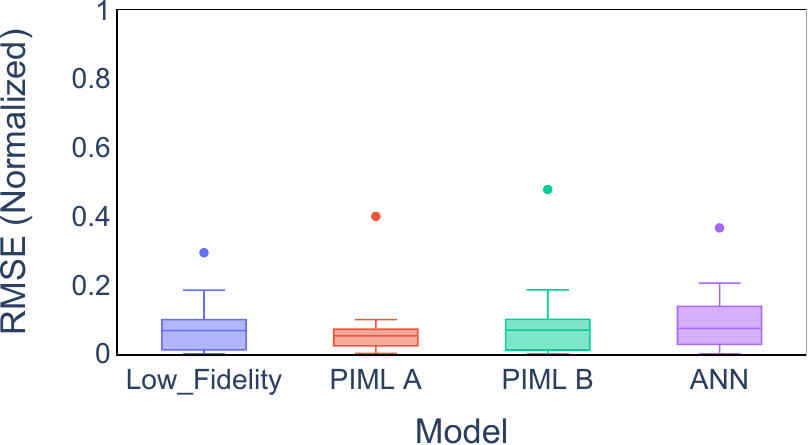}
        \caption{Prediction Error in $C_m$}
    \end{subfigure}
    \caption{Prediction Error Compared to High Fidelity Data }
    \label{fig:rmse}
\end{figure}

\subsection{Modelling Performance}

Figure.\ref{fig:ConvHist} shows the convergence history for both the PIML models. From the figure we see that PIML-A has a much smoother training history as compared to the PIML-B model. The PIML-B model seems to be sensitive to corrections in induced velocity. Further training with a lower learning rate or decaying the learning rate may resolve this issue and improve performance of the PIML-B model. It may also be the case that the VLM is very sensitive to the induced velocity inputs or that there is an unknown bias in the validation dataset. This could've happened due to the random splitting of data into train and test samples. Further work with varying test and train datasets similar to K-fold cross-validation are necessary to understand if there's any data bias.

\begin{figure}
    \begin{minipage}{0.49\linewidth}
        \centering
        \includegraphics[width=\linewidth]{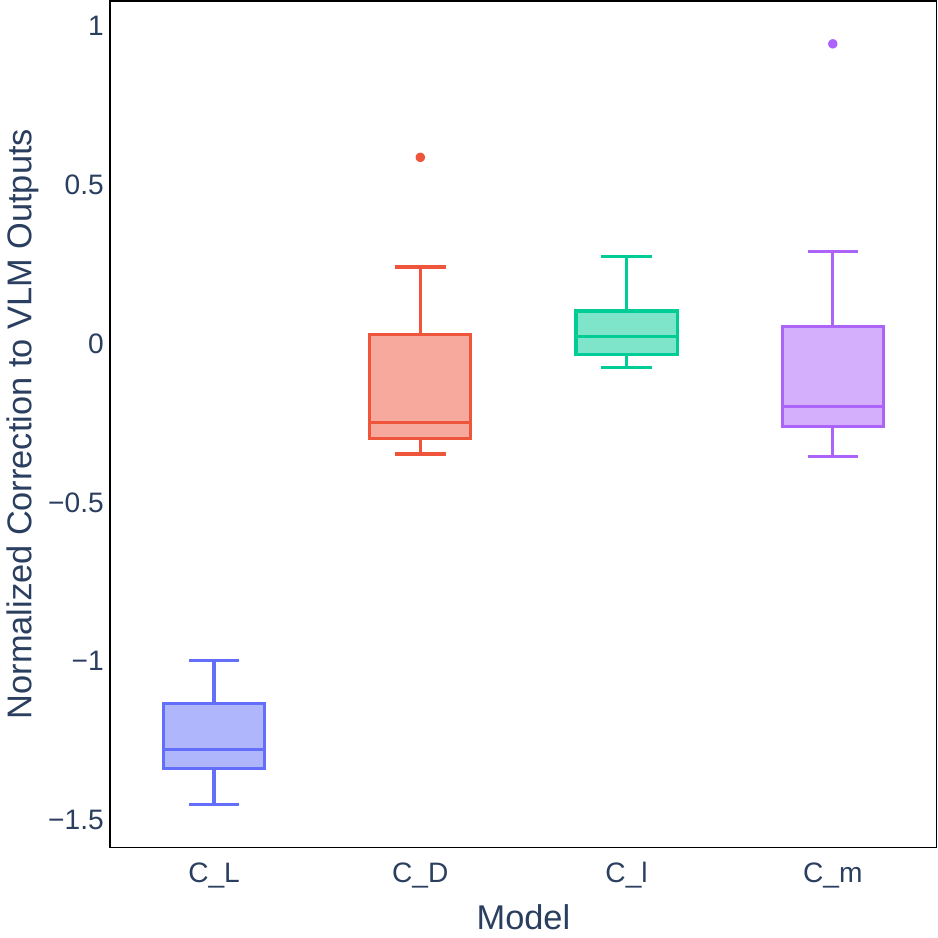}
        \captionof{figure}{Correction of VLM Outputs by the Correction Layers in PIML-A}
        \label{fig:correction_A}
    \end{minipage} \hfill 
    \begin{minipage}{0.49\linewidth}
        \centering
        \includegraphics[width=\linewidth]{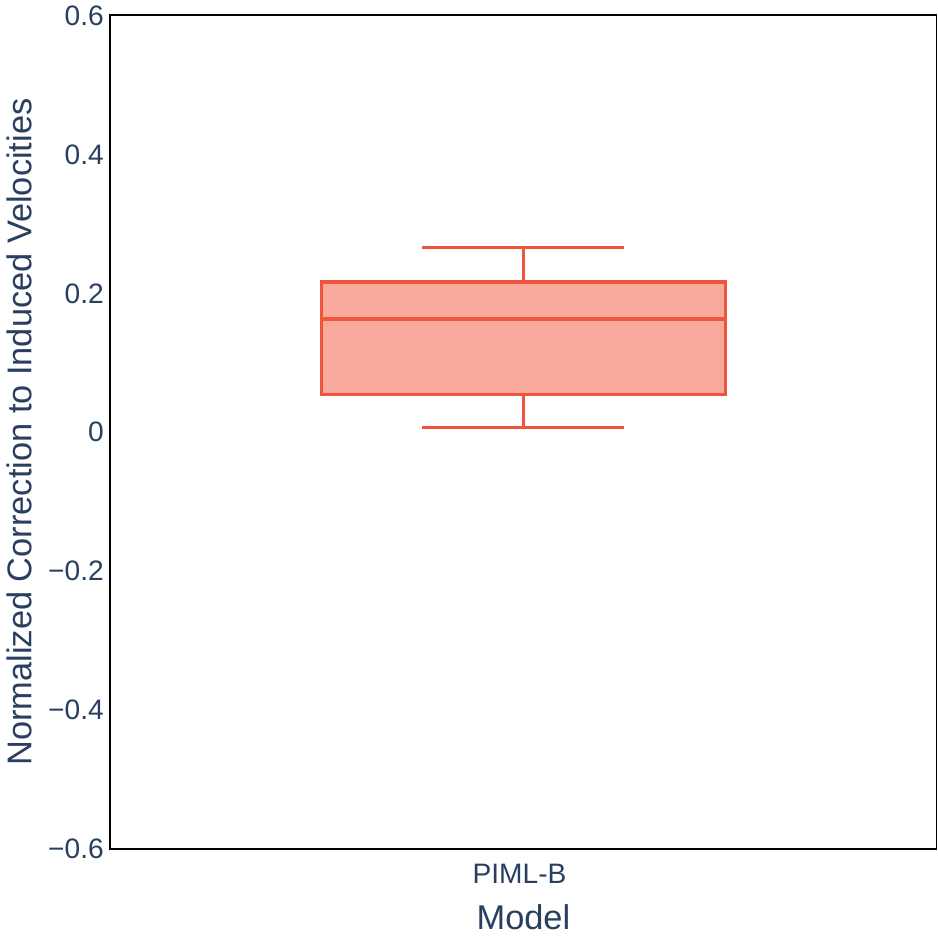}
        \captionof{figure}{Correction of Propeller Induced Velocities by the Correction Layers in PIML-B}
        \label{fig:correction_B}
    \end{minipage}
\end{figure}

Figure.\ref{fig:rmse} shows the prediction error of the models in all of the 4 outputs. From the figure, we see that the PIML-A model performs the best. The pure data-driven ANN and the PIML-B model exhibit similar performance. Relative to the amount of data available, the ANN performs very well and can prove to be a possible alternative in cases with a larger sample set and no additional requirements on interpretability. he fact that the high-fidelity samples were generated using Latin hypercube sampling, ensuring an even domain coverage, could also be a contributing factor to the ANN's strong performance. Among all the modeled outputs, PIML-A appears to excel in predicting $C_D$. This could be attributed to the transfer parameters having a greater impact on $C_D$ as compared to the other outputs. 
Where things begin to get interesting is when we see the effect of the learnt output correction layers in PIML-A as shown in fig.\ref{fig:correction_A}. From the figure, it is easily seen that the output layers of the VLM model have a very strong negative correction on the $C_L$ that the VLM outputs. As we also see from fig.\ref{fig:rmse}, we see that the VLM model has the highest prediction error in $C_L$. Combined, these two plots convey that the transfer parameters do not influence $C_L$ as much as they do the other outputs or that just shifting the inputs is not sufficient to improve the $C_L$ predictions of the VLM model. The correction for other outputs predicted by the PIML model is evenly distributed between positive and negative and this seems to be reasonable given the performance of the vanilla low-fidelity models on all these outputs is about the same.

The PIML-B model on the other hand performs on par with the low-fidelity VLM model. This outcome might be explained by an earlier hypothesis suggesting that if the induced velocities indeed have a minor impact on $C_L$, then regardless of what the correction network learns, it won't be able to enhance the predictions for $C_L$. However, since the training loss is calculated across all outputs, the overall loss remains substantial, causing the network to struggle in learning the transfer mapping necessary to improve predictions on the other inputs. It may be the case that when different models are trained to learn the various outputs of this modelling problem we might see improved performance by the PIML-B architecture on outputs other than $C_L$, but as that is not feasible especially once the number of outputs gets large enough to support design / control frameworks, PIML-A model might be better suited for this problem.

\section{Conclusion}
\label{sec:conclusion}

The paper presents two distinct architectures of physics-informed models, referred to as PIML-A and PIML-B. These models integrate Artificial Neural Networks with a low-fidelity BLOFI model with the aim of enhancing the prediction performance of the aerodynamic coefficients of a six-rotor eVTOL aircraft. To facilitate the training of both models, an end-to-end auto-differentiable physics framework was established using Google JAX. The models were then trained using a sparse dataset derived from a high-fidelity CHARM model. When comparing prediction performance, it was found that the PIML-A model achieved a higher degree of accuracy in comparison to the PIML-B model, the basic low-fidelity model, and a baseline purely data-driven model. A comprehensive analysis of the results demonstrated that the PIML-A model's method of simultaneous input-shifting and output-correction yielded superior results, given that one of the inputs might not be highly sensitive to the chosen transfer parameters. To conclusively determine the optimal architecture for this modeling problem, additional testing using cross-validation and hyperparameter optimization is needed. However, the initial results indicate that the PIML-A model holds significant promise. Looking forward, the BLOFI low-fidelity model, when augmented with PIML, could potentially be employed to generalize across diverse aircraft designs and configurations, thereby proving to be a valuable tool in the domains of aircraft design and control.

\section*{Acknowledgements}
This material is based upon work funded by Bechamo LLC's (with sub-contract to University at Buffalo) NASA Phase II \textit{Small Business Innovation Research (SBIR)} Award No. 80NSSC22CA046. The authors would also like to thank Flighthouse Engineering LLC for providing the aircraft design and the high-fidelity data used in this paper.
\printbibliography

\pagebreak
\appendix
\section{BLOFI Validation}
\begin{figure}[!h]
    \centering
    \begin{subfigure}{0.49\linewidth}
        \includegraphics[width = \linewidth]{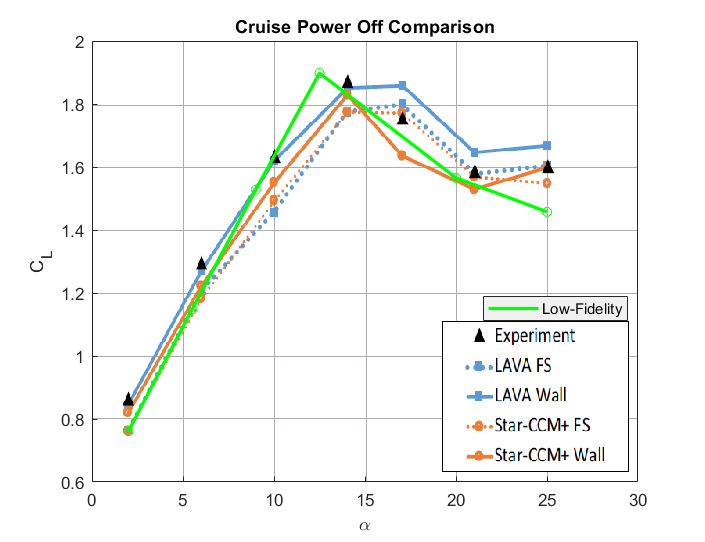}
    \end{subfigure} \hfill 
    \begin{subfigure}{0.49\linewidth}
        \includegraphics[width = \linewidth]{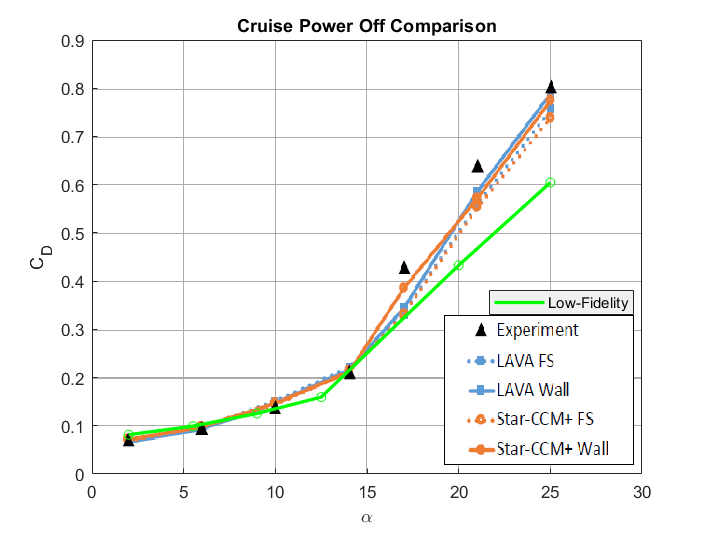}
    \end{subfigure} \\ 
    \begin{subfigure}{0.49\linewidth}
        \includegraphics[width = \linewidth]{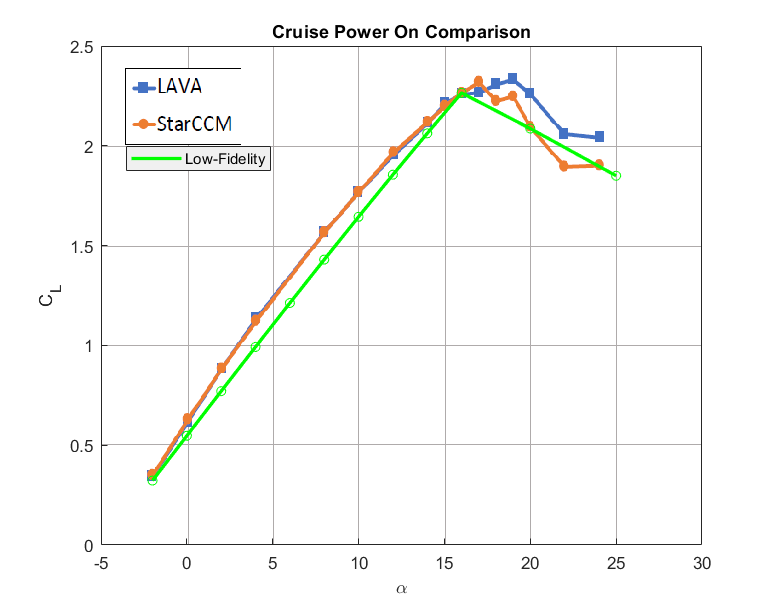}
    \end{subfigure} \hfill 
    \begin{subfigure}{0.49\linewidth}
        \includegraphics[width = \linewidth]{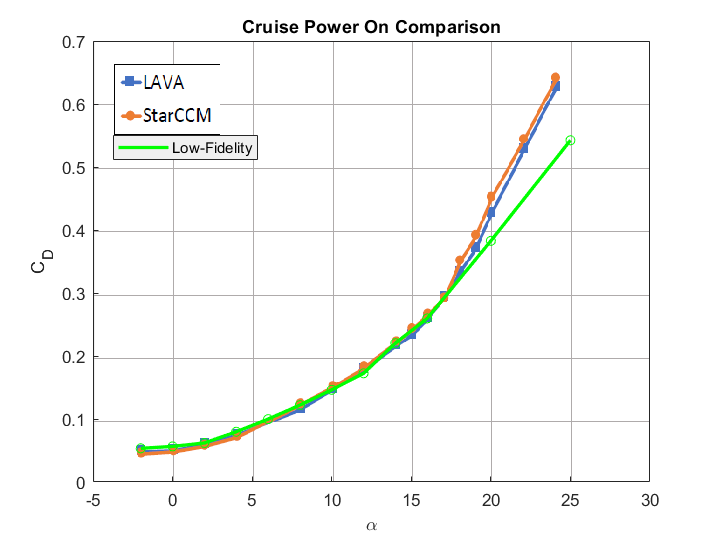}
    \end{subfigure} \\
    \begin{subfigure}{0.49\linewidth}
        \includegraphics[width = \linewidth]{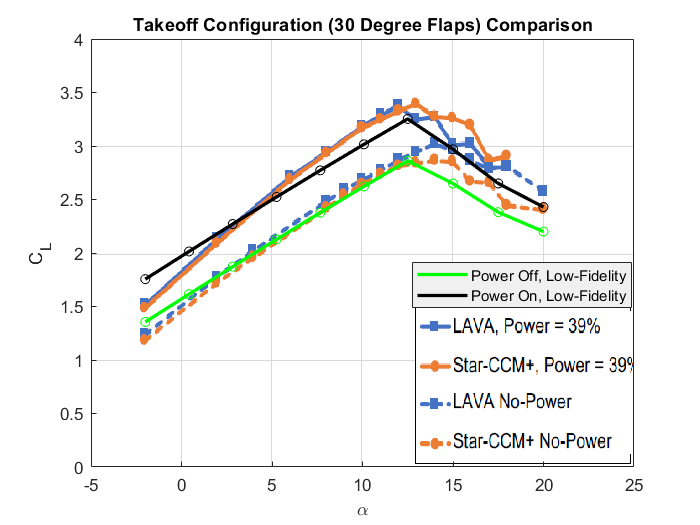}
    \end{subfigure} \hfill 
    \begin{subfigure}{0.49\linewidth}
        \includegraphics[width = \linewidth]{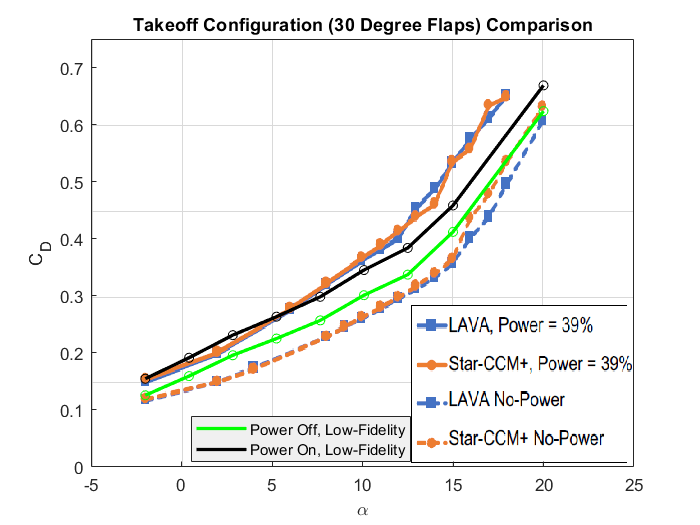}
    \end{subfigure}
    \caption{BLOFI was verified against X-57 CFD data.}
    \label{fig:X-57 vaidation}
\end{figure}

\end{document}